\documentclass[twocolumn,epjc3]{svjour3}          % twocolumn

\RequirePackage[T1]{fontenc}

\smartqed  % flush right qed marks, e.g. at end of proof

\RequirePackage{graphicx}
\RequirePackage{mathptmx}      % use Times fonts if available on your TeX system
\RequirePackage{flushend}
\RequirePackage[numbers,sort&compress]{natbib}
\RequirePackage[colorlinks,citecolor=blue,urlcolor=blue,linkcolor=blue]{hyperref}

\journalname{Eur. Phys. J. C}

\usepackage{graphics}

\newcommand{\p}{\partial}

\begin{document}
\title{Approximate Noether gauge symmetries of Bardeen model}
\author{U. Camci \thanksref{e1,addr1}}   % Do not remove
\thankstext{e1}{e-mail: ucamci@akdeniz.edu.tr}
\institute{Department of Physics, Faculty of Science, Akdeniz University, 07058 Antalya, Turkey \label{addr1} }
\date{Received: date / Revised version: date}

\maketitle

\begin{abstract}
We investigate the approximate Noether gauge symmetries of the geodesic Lagrangian for the Bardeen spacetime model. This is accommodated by a set of new approximate Noether gauge symmetry relations for the perturbed geodesic Lagrangian in the spacetime. A detailed analysis to the spacetime of Bardeen model up to third-order approximate Noether gauge symmetries is presented.
\end{abstract}

\section{Introduction}
\label{intro}
The black holes predicted in the framework of Einstein's gravity theory appear to exist in the Universe \cite{ford}. The existence of singularities inside the black holes was predicted by singularity theorems, and they indicate breakdown of the classical theory \cite{hawking}. A regular (i.e. non-singular) black hole which is characterized by the fact that the spacetime metric as well as the curvature invariants $R, R_{ab}$,$ R^{ab}$, $R_{abcd} R^{abcd}$ do not present singularities anywhere was firstly derived from Einstein's field equations by Bardeen \cite{bardeen} which was initially approximately obtained. As it was shown in Ref.\cite{garcia}, the metric of Bardeen model solves Einstein equations with a self-gravitating nonlinear magnetic monopole as a source. The Bardeen model includes two parameters in the solution, namely the parameters $m$ and $e$. As usual from the Schwarzschild metric, the parameter $m$ can be interpreted as the mass parameter for the system. The other parameter $e$ can be interpreted as the magnetic charge of the geometry. Although Bardeen model is always a regular black hole at $r=0$ for $e\neq 0$, it represents a regular black hole spacetime only when the inequality $27 e^2 \leq 16 m^2$ holds.

Let us briefly give some information about spacetime symmetries. A conformal Killing vector (CKV) ${\bf K}$ have to satisfy $\pounds_{\bf K} g_{ab} = 2 \psi (x^a) g_{ab}$, where $g_{ab}$ is the metric tensor,  $\pounds_{\bf K}$ is the Lie derivative operator along ${\bf K}$ and $\psi (x^a)$ is a conformal factor. When $\psi_{;ab} \neq 0$, the CKV field is said to be {\it proper} \cite{katzin}. The vector field ${\bf K}$ is called the special conformal Killing vector (SCKV) field if $\psi_{;ab} = 0$, the homothetic Killing vector (HKV) field if $\psi_{, \, a} = 0$, e.g. $\psi$ is a constant on the manifold, and the Killing vector (KV) field which gives the isometry if $\psi =0$. The set of all CKV (respectively SCKV, HKV and KV) form a finite-dimensional Lie algebra. The maximum dimension of the CKV algebra on the manifold $M$ is fifteen if $M$ is conformally flat, and it is seven if the spacetime is not conformally flat.

If a Lagrangian $L$ for a given dynamical system admits some symmetry, this property should strongly be related with Noether symmetries which describe physical features of differential equations possessing a Lagrangian $L$ in terms of first integrals admitted by them \cite{stephani,ibrahim}. This can actually be seen in two ways. First of all, one can consider a strict Noether symmetry approach \cite{capo00,camci2,camci3} which yields $\pounds_{\bf X} L = 0$, where $\pounds_{\bf X}$ is the Lie derivative operator along ${\bf X}$. On the other side, one could use the Noether gauge symmetry (NGS) approach \cite{feroze1,tsamparlis,camci1,ibrar,feroze2,camci2014a,camci2014b} which is a generalization of the strict Noether symmetry approach in the sense that the Noether symmetry equation includes a gauge term in such a way that the vector field ${\bf X} = \xi \p_s + \eta^a \frac{\p}{\p x^a}$ is called a NGS of a Lagrangian $L$ if there exists a gauge function, $A (s,x^a)$, such that the Noether symmetry condition holds \cite{stephani,ibrahim}
\begin{equation}\label{noether1}
{\bf X}^{[1]} L + L \, (D_{s}\xi) = D_{s} A ,
\end{equation}
where ${\bf X^{[1]}} = {\bf X}+\eta^a_{s} \frac{\partial}{\partial
\dot{x^a}}$ is the first prolongation operator of ${\bf X}$, ${\eta^a_{s}} = D_{s}\eta^a - \dot{x}^a D_{s}\xi$, and ${D_{s}} =  \frac{\partial}{\partial s} + \dot{x}^a \frac{\partial}{\partial x^a}$ is the total derivative operator. The approximate symmetries are as useful as exact ones in a variety of problems arising from physics and applied mathematics. We will discuss how the approximate Noether gauge symmetries (ANGSs) of the perturbed geodesic Lagrangian are related to the geometrical symmetries of spacetimes in the following section.

The geodesic Lagrangian describing the motion of the massive or massless particles under the influence of a potential function $V(x^c)$ is
\begin{equation}
L(s, x^c, \dot{x}^c) = \frac{1}{2} g_{ab} \dot{x}^a \dot{x}^b - V(x^c),  \label{lagr}
\end{equation}
where the dot represents derivative with respect to the geodetic parameter $s$. The significance of an ANGS comes from the fact that a manifold having no exact symmetry can possess approximate ones. In this context, using the approximate symmetries of perturbed geodesic Lagrangian for spacetimes, one can recover the \emph{lost} symmetries and conservation laws corresponding to the perturbed Lagrangian \cite{feroze3, bokhari2006, kara2008, hussain2007,hussain2009}.

For the Schwarzschild metric, Kara \emph{et al.}\cite{kara2008} have considered the approximate symmetries and the conservation laws of geodesic equations. Following the use of approximate symmetries for the Schwarzschild spacetime, Hussain \emph{et al.}\cite{hussain2007} have investigated the exact and approximate symmetries of the system of geodesic equations for the Reissner-Nordstr\"{o}m (RN) spacetime. Using the ANGS for the Schwarzschild and Kerr metrics, Hussain \emph{et al.} \cite{hussain2009} have recovered all the lost conservation laws as trivial first-order approximate conservation laws in first-order approximate case. Using approximate Lie symmetries the energy content of colliding plane waves have studied by Sharif and Waheed \cite{sharif2012}. By analyzing the behavior of the effective potential for the particles and photons, it is investigated the time-like and null geodesic structures for the Bardeen metric\cite{wang}. In a recent paper, Sharif and Waheed \cite{sharif2011} have applied the approximate Lie symmetry method to investigate the energy re-scaling factor for the Bardeen's regular black-hole solution. They only evaluated the third-order approximate symmetries of the orbital and geodesic equations using approximate Lie symmetry methods for differential equations. Our main aim here is to extend our mathematical understanding of equation (\ref{noether1}) by exploring its geometrical properties. We have obtained the geometrical nature of ANGSs up to second-order perturbed Lagrangian and applied this method to the Schwarzschild and RN spacetimes \cite{camci2014c}. It should be stressed that we deal with third-order ANGSs of the perturbed geodesic Lagrangian which are not studied previously as far as we know.

The rest of the paper is organized as follows. In the following section, we present an analysis of third-order ANGSs of first-order geodesic Lagrangian according to perturbed spacetime geometry. In section \ref{sec:2}, we apply the obtained results of section \ref{sec:1} to the spacetime of Bardeen model. Conclusions and discussions are presented in last section \ref{sec:4}.

\section{Third Order Approximate Noether gauge symmetries}
\label{sec:1}

Calculation method of exact symmetries and the first-order approximate symmetries of a Lagrangian are available in the literature \cite{feroze3,bokhari2006,kara2008}. Hussain \emph{et al.} \cite{hussain2007,hussain2009} have extended the procedure of calculating the approximate symmetries of a perturbed Lagrangian to the second-order. In this study we introduce a new geometrical method to calculate the ANGSs of the first-order perturbed Lagrangian extending the procedure of obtaining ANGSs until the third-order.

The third-order ANGSs of the first-order perturbed Lagrangian
\begin{eqnarray}
& & L(s, x^a, \dot{x}^a, \epsilon) = L_0 (s, x^a, \dot{x}^a) + \epsilon L_1 (s, x^a, \dot{x}^a) \nonumber \\& & \qquad \quad \qquad + \epsilon^2 L_2 (s, x^a, \dot{x}^a) + \epsilon^3 L_3 (s, x^a, \dot{x}^a) +  O(\epsilon^4), \label{p-lagr}
\end{eqnarray}
is given by the ANGS generator
\begin{equation}
{\bf X} = {\bf X}_0 + \epsilon {\bf X}_1 + \epsilon^2 {\bf X}_2 + \epsilon^3 {\bf X}_3,
\end{equation}
up to the gauge function $A (s,x^a) = A_0 (s,x^a) + \epsilon A_1 (s,x^a) + \epsilon^2 A_2 (s,x^a) + \epsilon^3 A_3 (s,x^a)$ if the ANGS generator satisfies the approximate Noether gauge symmetry conditions
\begin{eqnarray}
& & {\bf X}_0^{[1]} L_0 + L_0 \, (D_{s}\xi_0) = D_{s} A_0 , \label{noet1} \\& & {\bf X}_1^{[1]} L_0 + {\bf X}_0^{[1]} L_1 + L_0 \, (D_{s}\xi_1) + L_1 \, (D_{s}\xi_0)= D_{s} A_1, \label{noet2} \\ & & {\bf X}_2^{[1]} L_0 + {\bf X}_1^{[1]} L_1 + {\bf X}_0^{[1]} L_2 +  L_0 \, (D_{s}\xi_2) + L_1 \, (D_{s}\xi_1) \nonumber \\ & & \qquad  +  L_2 \, (D_{s}\xi_0) = D_{s} A_2, \label{noet3}\\ & & {\bf X}_3^{[1]} L_0 + {\bf X}_2^{[1]} L_1 + {\bf X}_1^{[1]} L_2 + {\bf X}_0^{[1]} L_3 + L_0 \, (D_{s}\xi_3) \nonumber \\ & & \qquad  + L_1 \, (D_{s}\xi_2) +  L_2 \, (D_{s}\xi_1) + L_3 \, (D_{s}\xi_0) = D_{s} A_3, \label{noet4}
\end{eqnarray}
where  ${\bf X}_0$ is the exact NGS generator,  ${\bf X}_1, {\bf X}_2$ and ${\bf X}_3$ are the first-order, second-order and third-order ANGS generators, respectively, which are defined as
\begin{eqnarray}
& & {\bf X}_j = \xi_j \frac{\p}{\p s} + \eta^a_j \frac{\p}{\p x^a}, \quad (j = 0,1,2,3), \\ & & {\bf X}_j^{[1]} = {\bf X}_j+\eta^a_{j(s)} \frac{\partial}{\partial
\dot{x}^a}, \quad {\eta^a_{j(s)}} = D_{s}\eta^a_j - \dot{x}^a D_{s}\xi_j,
\end{eqnarray}
The above perturbed Lagrangian (\ref{p-lagr}) yields a third order (in $\epsilon$) perturbed system of ordinary differential equations (ODEs). If ${\bf X}_j$ are the ANGSs corresponding to the perturbed Lagrangians $L_j (s, x^a, \dot{x}^a)$, then
\begin{eqnarray}
& & I_0 = \xi_0 L_0 + \left(\eta^a_0 - \xi_0 \dot{x}^a \right) \frac{\partial
L_0}{\partial \dot{x}^a} - A_0, \label{fint-X0} \\ & & I_1 = \xi_0 L_1 + \xi_1 L_0 + \left(\eta^a_0 - \xi_0 \dot{x}^a \right) \frac{\partial L_1}{\partial \dot{x}^a} \nonumber \\& & \qquad + \left(\eta^a_1 - \xi_1 \dot{x}^a \right) \frac{\partial L_0}{\partial \dot{x}^a} - A_1, \label{fint-X1} \\
& & I_2 =  \xi_0 L_2 + \xi_1 L_1 + \xi_2 L_0 + \left(\eta^a_0 - \xi_0 \dot{x}^a \right) \frac{\partial L_2}{\partial \dot{x}^a}  \nonumber \\& & \qquad + \left(\eta^a_1 - \xi_1 \dot{x}^a \right) \frac{\partial L_1}{\partial \dot{x}^a} + \left(\eta^a_2 - \xi_2 \dot{x}^a \right) \frac{\partial L_0}{\partial \dot{x}^a} - A_2, \label{fint-X2}\\
& & I_3 =  \xi_0 L_3 + \xi_1 L_2 + \xi_2 L_1 +  \xi_3 L_0 + \left(\eta^a_0 - \xi_0 \dot{x}^a \right) \frac{\partial L_3}{\partial \dot{x}^a}  \nonumber \\& & \qquad + \left(\eta^a_1 - \xi_1 \dot{x}^a \right) \frac{\partial L_2}{\partial \dot{x}^a} + \left(\eta^a_2 - \xi_2 \dot{x}^a \right) \frac{\partial L_1}{\partial \dot{x}^a} \nonumber \\& & \qquad + \left(\eta^a_3 - \xi_3 \dot{x}^a \right) \frac{\partial L_0}{\partial \dot{x}^a} - A_3, \label{fint-X3}
\end{eqnarray}
are the first integrals associated with ANGSs ${\bf X}_j$ (j=0,1,2,3).

The spacetime metric $g_{ab}$ can be decomposed as follows:
\begin{equation}
g_{ab} = \gamma_{ab} + \epsilon h_{ab} + \epsilon^2 \sigma_{ab} + \epsilon^3 k_{ab}, \label{p-metric}
\end{equation}
which means by (\ref{lagr}) and (\ref{p-lagr}) that the exact and perturbed geodesic Lagrangians of motion have the form
\begin{eqnarray}
& & L_0 (s, x^a, \dot{x}^a) = \frac{1}{2} \gamma_{ab} \dot{x}^a \dot{x}^b - V(x^c),   \label{lagr0} \\& & L_1 (s, x^a, \dot{x}^a) = \frac{1}{2} h_{ab} \dot{x}^a \dot{x}^b, \label{lagr1} \\& &  L_2 (s, x^a, \dot{x}^a) = \frac{1}{2} \sigma_{ab} \dot{x}^a \dot{x}^b, \label{lagr2}\\& &  L_3 (s, x^a, \dot{x}^a) = \frac{1}{2} k_{ab} \dot{x}^a \dot{x}^b, \label{lagr3}
\end{eqnarray}
where $\gamma_{ab}, h_{ab}, \sigma_{ab}$ and $k_{ab}$ are the exact, the first-order, the second-order and the third-order perturbed metrics, respectively. The metric $\gamma_{ab}$ should be non-degenerate (i.e., $\det (\gamma_{ab}) \neq 0$). But the other metrics  $h_{ab}, \sigma_{ab}$ and $k_{ab}$ can be degenerate (i.e., $\det (h_{ab}) = 0, \det (\sigma_{ab}) = 0, \det (k_{ab}) = 0$) or non-degenerate, and they represent the slight deviations from flat spacetime geometry if the metric $\gamma_{ab}$ represents flat geometry.

For the ANGS conditions (\ref{noet1}) - (\ref{noet4}) of the perturbed geodesic Lagrangians (\ref{lagr0}) - (\ref{lagr3}) we obtain
\begin{eqnarray}
& & {\bf X}_j^{[1]} L_0 =  (\gamma_{ad} \eta^d_{j,s}) \dot{x}^a + \frac{1}{2} \left( \pounds_{\bf \eta_j} \gamma_{ab} -2 \xi_{j,s} \gamma_{ab} \right) \dot{x}^a \dot{x}^b  \nonumber\\ & &  \qquad \qquad - (\gamma_{bd} \xi_{j,a}) \dot{x}^a \dot{x}^b \dot{x}^d - \eta^d_j V_{,d} ,   \\ & & {\bf X}_j^{[1]} L_1 = (h_{ad} \eta^d_{j,s}) \dot{x}^a + \frac{1}{2} \left( \pounds_{\bf \eta_j} h_{ab} -2 \xi_{j,s} h_{ab} \right) \dot{x}^a \dot{x}^b \nonumber\\ & &  \qquad \qquad - (h_{bd} \xi_{j,a}) \dot{x}^a \dot{x}^b \dot{x}^d, \quad  \\ & & {\bf X}_j^{[1]} L_2 = (\sigma_{ad} \eta^d_{j,s}) \dot{x}^a + \frac{1}{2} \left( \pounds_{\bf \eta_j} \sigma_{ab} -2 \xi_{j,s} \sigma_{ab} \right) \dot{x}^a \dot{x}^b \nonumber\\ & &  \qquad \qquad - (\sigma_{bd} \xi_{j,a}) \dot{x}^a \dot{x}^b \dot{x}^d, \quad \\ & & {\bf X}_j^{[1]} L_3 = (k_{ad} \eta^d_{j,s}) \dot{x}^a + \frac{1}{2} \left( \pounds_{\bf \eta_j} k_{ab} -2 \xi_{j,s} k_{ab} \right) \dot{x}^a \dot{x}^b \nonumber\\ & &  \qquad \qquad - (k_{bd} \xi_{j,a}) \dot{x}^a \dot{x}^b \dot{x}^d, \quad
\end{eqnarray}
where $\pounds_{\bf \eta_0}, \pounds_{\bf \eta_1}, \pounds_{\bf \eta_2}$ and $\pounds_{\bf \eta_3}$ are the geometrical derivative, i.e. Lie derivative, operators along ${\bf \eta_0} = \eta^a_0 \p_{x^a}$, ${\bf \eta_1} = \eta^a_1 \p_{x^a}$, ${\bf \eta_2} = \eta^a_2 \p_{x^a}$ and ${\bf \eta}_3 = \eta^a_3 \p_{x^a}$, respectively. Putting the above expressions into (\ref{noet1})-(\ref{noet4}) together with $D_s \xi_j = \xi_{j,s} + \xi_{j,d} \dot{x}^d$ and  $D_s A_j = A_{j,s} + A_{j,d} \dot{x}^d$, (j=0,1,2,3), we find the ANGS conditions for the Eq.(\ref{noet1})
\begin{eqnarray}
& & \xi_{0,a} = 0,  \,\,  \gamma_{ab} \eta^b_{0,s} = A_{0,a}  \label{aneq-1-1}  \\& &  \pounds_{\bf \eta_0} \gamma_{ab} = \xi_{0,s} \gamma_{ab} \label{aneq-1-2} \\& &  \pounds_{\bf \eta_0} V = - \xi_{0,s} V - A_{0,s} \label{aneq-1-3}
\end{eqnarray}
for the Eq.(\ref{noet2})
\begin{eqnarray}
& & \xi_{1,a} = 0,  \quad  \gamma_{ab} \eta^b_{1,s} + h_{ab} \eta^b_{0,s} = A_{1,a} \label{aneq-2-1} \\& & \pounds_{\bf \eta_1} \gamma_{ab} + \pounds_{\bf \eta_0} h_{ab} = \xi_{1,s} \gamma_{ab} + \xi_{0,s} h_{ab} \label{aneq-2-2} \\& &  \pounds_{\bf \eta_1} V = - \xi_{1,s} V - A_{1,s} \label{aneq-2-3}
\end{eqnarray}
for the Eq.(\ref{noet3})
\begin{eqnarray}
& & \xi_{2,a} = 0,  \quad  \gamma_{ab} \eta^b_{2,s} + h_{ab} \eta^b_{1,s} + \sigma_{ab} \eta^b_{0,s} = A_{2,a} \label{aneq-3-1}\\& & \pounds_{\bf \eta_2} \gamma_{ab} + \pounds_{\bf \eta_1} h_{ab}  + \pounds_{\bf \eta_0} \sigma_{ab} \nonumber \\& & \qquad \quad = \xi_{2,s} \gamma_{ab} + \xi_{1,s} h_{ab} + \xi_{0,s} \sigma_{ab} \label{aneq-3-2} \\& &  \pounds_{\bf \eta_2} V = - \xi_{2,s} V - A_{2,s} \,\, \label{aneq-3-3}
\end{eqnarray}
and for the Eq.(\ref{noet4})
\begin{eqnarray}
& & \xi_{3,a} = 0,  \gamma_{ab} \eta^b_{3,s} + h_{ab} \eta^b_{2,s} + \sigma_{ab} \eta^b_{1,s} + k_{ab} \eta^b_{0,s} = A_{3,a} \qquad \label{aneq-4-1}\\& & \pounds_{\bf \eta_3} \gamma_{ab} + \pounds_{\bf \eta_2} h_{ab} + \pounds_{\bf \eta_1} \sigma_{ab} + \pounds_{\bf \eta_0} k_{ab} \nonumber \\& & \qquad \quad  = \xi_{3,s} \gamma_{ab} + \xi_{2,s} h_{ab} + \xi_{1,s} \sigma_{ab} + \xi_{0,s} k_{ab}\label{aneq-4-2} \\& &  \pounds_{\bf \eta_3} V = - \xi_{3,s} V - A_{3,s} \ . \label{aneq-4-3}
\end{eqnarray}
Thus we find the geometrical form of ANGS equations in terms of Lie derivatives of exact and perturbed metrics up to the third-order. The Equations (\ref{aneq-1-1}) - (\ref{aneq-1-3}) are, of course, the conditions used to determine NGSs of exact geodesic Lagrangian. We have shown that the exact (\ref{noet1}), the first-order (\ref{noet2}), the second-order (\ref{noet3}) and the third-order (\ref{noet4}) form of ANGS conditions are equivalent to the set of equations (\ref{aneq-1-1}) - (\ref{aneq-1-3}), (\ref{aneq-2-1})-(\ref{aneq-2-3}), (\ref{aneq-3-1})-(\ref{aneq-3-3}) and (\ref{aneq-4-1})-(\ref{aneq-4-3}), respectively.

\section{Application to Bardeen Model}
\label{sec:2}

The Bardeen model describes a singularity-free black hole spacetime which means it is a regular spacetime. In this section, we will apply the above geometrical method to the third-order perturbed spacetime of Bardeen model for finding ANGSs of perturbed geodesic Lagrangian. The line element representing the Bardeen spacetime is given by
\begin{eqnarray}\label{bardeen-metric}
& & ds^2 = \left[ 1 - \frac{2 m r^2}{(r^2 + e^2)^{3/2}} \right] dt^2 - \left[ 1 - \frac{2 m r^2}{(r^2 + e^2)^{3/2}} \right]^{-1} dr^2 \nonumber \\& & \quad \qquad - r^2 \left( d\theta^2 + \sin^2 \theta d\phi^2 \right),
\end{eqnarray}
where $m$ is the mass of configuration and $e$ represents the magnetic charge of the nonlinear self-gravitating monopole, and we have taken the gravitational units, i.e. $G = c =1$. The Bardeen black hole metric reduces to the Schwarzschild metric for  $e= 0$ and is flat for $m = 0$.
Using the expansions
\begin{equation}\label{Bardeen-exp1}
g_{tt} = 1 - \frac{\epsilon}{r} + \frac{3 k \epsilon^3}{2 r^3} + O (\epsilon^4),
\end{equation}
and
\begin{equation}\label{Bardeen-exp2}
g_{rr} = -\left[ 1 + \frac{\epsilon}{r} + \frac{\epsilon^2}{r^2} + \left( 1 - \frac{3 k}{2} \right) \frac{\epsilon^3}{r^3} \right] + O (\epsilon^4),
\end{equation}
the first-order perturbed geodesic Lagrangian of Bardeen model is given by
\begin{eqnarray}
& & L = \frac{1}{2} \left[ \dot{t}^2 - \dot{r}^2 - r^2 (\dot{\theta}^2 + \sin^2 \theta \, \dot{\phi}^2 ) \right]  - \frac{\epsilon}{2 r} \left(  \dot{t}^2 + \dot{r}^2 \right) \nonumber \\& & \qquad - \frac{\epsilon^2}{2 r^2} \dot{r}^2  + \frac{\epsilon^3}{2 r^3} \left[ \frac{3k}{2} \dot{t}^2 - \left( 1- \frac{3k}{2}  \right) \dot{r}^2  \right] \nonumber \\& & \qquad - V(t,r,\theta,\phi) + O(\epsilon^4), \label{Bardeen-lagr}
\end{eqnarray}
where we take $\epsilon = 2m$ and $e^2 = k \epsilon^2$, $0 < k \leq 4/27$, in the limit of small mass of a point gravitating source. The above Lagrangian (\ref{Bardeen-lagr}) yields
\begin{eqnarray}
& & L_0 = \frac{1}{2} \left[ \dot{t}^2 - \dot{r}^2 - r^2 (\dot{\theta}^2 + \sin^2 \theta \, \dot{\phi}^2 ) \right], \label{Bardeen-lagr0} \\& & L_1 = - \frac{1}{2 r} \left(  \dot{t}^2 + \dot{r}^2 \right), \label{Bardeen-lagr1}\\& & L_2 = - \frac{1}{2 r^2} \dot{r}^2 , \label{Bardeen-lagr2}\\& & L_3 =  \frac{1}{2 r^3} \left[ \frac{3 k}{2} \dot{t}^2 - \left( 1 - \frac{3k}{2}\right) \dot{r}^2 \right), \label{Bardeen-lagr3}
\end{eqnarray}
where we used $\gamma_{ab} = {\rm diag} (1, -1, -r^2, -r^2 \sin^2 \theta)$ which is the Minkowski metric,  $h_{ab} = {\rm diag}(-1/ r, -1/ r, 0,0)$, $\sigma_{ab} = {\rm diag}(-1/r^2,0,0,0)$ and $k_{ab} = {\rm diag}(3k/ 2 r^3, (3k/2 -1)/r^3,0,0)$. Also, the Lagrangian (\ref{Bardeen-lagr}) reduces to the geodesic Lagrangian of the Minkowski metric in the limit $\epsilon \rightarrow 0$. Thus, we can apply the new set of ANGSs obtained in the previous section to these exact metric $\gamma_{ab}$ and perturbed metrics  $h_{ab}, \sigma_{ab}$ and $k_{ab}$ for the Bardeen spacetime. For the metric $\gamma_{ab}$, we find from the exact ANGS Eqs. (\ref{aneq-1-1})-(\ref{aneq-1-3}) the following 19 equations
\begin{eqnarray}
& & \xi_{0,t} = 0, \quad \xi_{0,r} = 0, \quad \xi_{0,\theta} = 0, \quad \xi_{0,\phi} = 0, \nonumber \\& & \eta^0_{0,s} - A_{0,t} = 0, \,\, \eta^1_{0,s} + A_{0,r} = 0, \,\, r^2 \eta^2_{0,s} + A_{0,\theta} = 0, \nonumber \\& & r^2 \sin^2 \theta \eta^3_{0,s} + A_{0,\phi} = 0, \quad 2 \eta^0_{0,t} - \xi_{0,s} = 0, \nonumber \\& & 2 \eta^1_{0,r} - \xi_{0,s} = 0, \quad 2 \eta^2_{0,\theta} + \frac{2}{r} \eta^1_0 - \xi_{0,s} = 0, \nonumber \\& & \frac{2}{r} \eta^1_0 + 2 \cot \theta \eta^2_0 + 2 \eta^3_{0,\phi} - \xi_{0,s} = 0,  \label{angs-eq1} \\& & \eta^0_{0,r} - \eta^1_{0,t} = 0, \,\, \eta^0_{0,\theta} - r^2 \eta^2_{0,t} = 0, \nonumber \\& & \eta^0_{0,\phi} - r^2 \sin^2 \theta \, \eta^3_{0,t} = 0, \,\, \eta^1_{0,\theta} + r^2 \eta^2_{0,r} = 0, \nonumber \\& & \eta^1_{0,\phi} + r^2 \sin^2 \theta \, \eta^3_{0,r} = 0, \,\, \eta^2_{0,\phi} + \sin^2 \theta \eta^3_{0,\theta} = 0, \nonumber \\& & V_{,t} \eta^0_0 + V_{,r} \eta^1_0 + V_{,\theta} \eta^2_0 + V_{,\phi} \eta^3_0 + V \xi_{0,s} + A_{0,s} = 0. \nonumber
\end{eqnarray}
Considering both $\gamma_{ab}$ and $h_{ab}$, the first-order ANGS Eqs.(\ref{aneq-2-1})-(\ref{aneq-2-3}) yield again 19 equations
\begin{eqnarray}
& & \xi_{1,t} = 0, \quad \xi_{1,r} = 0, \quad \xi_{1,\theta} = 0, \quad \xi_{1,\phi} = 0, \nonumber \\& & \eta^0_{1,s} - \frac{1}{r} \eta^1_{0,s} - A_{1,t} = 0, \,\, \eta^1_{1,s} + \frac{1}{r} \eta^1_{0,s} + A_{1,r} = 0, \nonumber \\& &  r^2 \eta^2_{1,s} + A_{1,\theta} = 0, \,\, r^2 \sin^2 \theta \eta^3_{1,s} + A_{1,\phi} = 0, \nonumber \\& & \frac{1}{r} \eta^1_0 - 2 \eta^0_{0,t} + 2 r \eta^0_{1,t} + \xi_{0,s}  - r \xi_{1,s} = 0, \nonumber \\& & \frac{1}{r} \eta^1_0 - 2 \eta^1_{0,r} - 2 r \eta^1_{1,r} + \xi_{0,s} + \xi_{1,s} = 0,  \nonumber \\& & \frac{2}{r} \eta^1_1 + 2 \cot \theta \eta^2_1 + 2 \eta^3_{1,\phi} - \xi_{1,s} = 0, \label{angs-eq2} \\& & \frac{2}{r} \eta^1_1 + \eta^2_{1,\theta} - \xi_{1,s} = 0, \,\, \eta^0_{1,r} - \frac{1}{r} \left( \eta^0_{0,r} + \eta^1_{0,t} \right) -\eta^1_{1,t} = 0, \nonumber \\& &  \eta^0_{1,\theta} - \frac{1}{r} \eta^0_{0,\theta} - r^2 \eta^2_{1,t} = 0, \quad \eta^1_{1,\theta} + \frac{1}{r} \eta^1_{0,\theta} + r^2 \eta^2_{1,r} = 0, \nonumber \\& &  \eta^0_{1,\phi} - \frac{1}{r} \eta^0_{0,\phi} - r^2 \sin^2 \theta \, \eta^3_{1,t} = 0, \nonumber \\& &  \eta^1_{1,\phi} + \frac{1}{r} \eta^1_{0,\phi} + r^2 \sin^2 \theta \, \eta^3_{1,r} = 0, \nonumber \\& &  \eta^2_{1,\phi} + \sin^2 \theta \eta^3_{1,\theta} = 0, \nonumber \\& &  V_{,t} \eta^0_1 + V_{,r} \eta^1_1 + V_{,\theta} \eta^2_1 + V_{,\phi} \eta^3_1 + V \xi_{1,s} + A_{1,s} = 0.  \nonumber
\end{eqnarray}
Using the metrics $\gamma_{ab}, h_{ab}$ and $\sigma_{ab}$, the second-order ANGS Eqs.(\ref{aneq-3-1})-(\ref{aneq-3-3}) give the following equations
\begin{eqnarray}
& & \xi_{2,t} = 0, \quad \xi_{2,r} = 0, \quad \xi_{2,\theta} = 0, \quad \xi_{2,\phi} = 0, \nonumber \\& & \frac{1}{r} \eta^0_{1,s} - \eta^0_{2,s} + A_{2,t} = 0, \nonumber \\& & \frac{1}{r^2} (r ~ \eta^1_{1,s} + \eta^1_{0,s}) + \eta^1_{2,s} + A_{2,r} = 0, \nonumber \\& &  r^2 \eta^2_{2,s} + A_{2,\theta} = 0, \,\,  r^2 \sin^2 \theta \eta^3_{2,s} + A_{2,\phi} = 0, \nonumber\\& & \frac{1}{r^2} (\eta^1_1 - 2 r ~\eta^0_{1,t} + r ~ \xi_{1,s} ) + 2 \eta^0_{2,t} - \xi_{2,s} = 0, \nonumber \\& & \frac{1}{r^3} ( 2 \eta^1_0 + r ~ \eta^1_1 - 2 r^2 \eta^1_{1,r} - 2 r^2 \eta^1_{0,r} +  r ~ \xi_{0,s} + r^2 \xi_{1,s}) \nonumber \\& & - 2 \eta^1_{2,r} + \xi_{2,s} = 0, \quad  \frac{2}{r} \eta^1_2 + 2 \eta^2_{2,\theta} - \xi_{2,s} = 0,  \label{angs-eq3} \\& & \frac{2}{r} \eta^1_2 + 2 \cot \theta \eta^2_2 + 2 \eta^3_{2,\phi} - \xi_{2,s} = 0, \,\, \eta^2_{2,\phi} + \sin^2 \theta \eta^3_{2,\phi} = 0, \nonumber \\& & \eta^0_{2,\theta} - r^2 \eta^2_{2,t} - \frac{1}{r} \eta^0_{1,\theta} = 0, \,\, \eta^0_{2,\phi} - r^2 \sin^2 \theta \eta^3_{2,t} - \frac{1}{r} \eta^0_{1,\phi} = 0, \nonumber \\& & \eta^1_{2,\theta} + r^2 \eta^2_{2,r} + \frac{1}{r} \eta^1_{1,\theta} + \frac{1}{r^2} \eta^1_{0,\theta} = 0, \nonumber \\& & \eta^1_{2,\phi} + r^2 \sin^2 \theta \eta^3_{2,r} + \frac{1}{r} \eta^1_{1,\phi} + \frac{1}{r^2} \eta^1_{0,\phi} = 0, \nonumber\\
& &  V_{,t} \eta^0_2 + V_{,r} \eta^1_2 + V_{,\theta} \eta^2_2 + V_{,\phi} \eta^3_2 + V \xi_{2,s} + A_{2,s} = 0.  \nonumber
\end{eqnarray}
Finally, the third-order ANGS Eqs. (\ref{aneq-4-1})-(\ref{aneq-4-3}) of the metrics $\gamma_{ab}, h_{ab}, \sigma_{ab}$ and $k_{ab}$ are explicitly written as
\begin{eqnarray}
& & \xi_{3,t} = 0, \quad \xi_{3,r} = 0, \quad \xi_{3,\theta} = 0, \quad \xi_{3,\phi} = 0, \nonumber \\& &  \eta^0_{3,s} - \frac{1}{r} \eta^0_{2,s} \frac{3 k}{2 r^3} \eta^0_{0,s}- A_{3,t} = 0, \nonumber \\& & \frac{1}{r^3} [ (1 - \frac{3 k}{2}) \eta^1_{0,s} + r ~ \eta^1_{1,s} + r^2 \eta^1_{2,s} ] + \eta^1_{3,s} + A_{3,r} = 0, \nonumber \\& &  r^2 \eta^2_{3,s} + A_{3,\theta} = 0, \,\,  r^2 \sin^2 \theta \eta^3_{3,s} + A_{3,\phi} = 0, \nonumber\\& & 2 \eta^0_{3,t} + \frac{1}{r^2} (\eta^1_2 - 2 r ~\eta^0_{2,t} + r ~ \xi_{2,s} ) \nonumber \\& & \quad - \frac{9 k}{2 r^4} \eta^1_0 + \frac{3k}{2 r^3} ( 2 \eta^0_{0,t} - \xi_{0,s}) - \xi_{3,s} = 0, \nonumber \\& & 2 \eta^1_1 +  r ~ \eta^1_2 - 2 r^2 \eta^1_{2,r} - 2 r \eta^1_{1,r} +  r ~ \xi_{1,s} + r^2 \xi_{2,s} - \frac{3k}{2} \xi_{0,s} \nonumber \\& &  \,\, + \frac{3 (3k-2)}{2 r} \eta^1_0 + (3k -2) \eta^1_{0,r} + r^3 (- 2 \eta^1_{3,r} + \xi_{3,s}) = 0, \nonumber \\& &  \frac{2}{r} \eta^1_3 + 2 \eta^2_{3,\theta} - \xi_{3,s} = 0,  \,\, \eta^2_{3,\phi} + \sin^2 \theta \eta^3_{3,\phi} = 0, \label{angs-eq4} \\& & \frac{2}{r} \eta^1_3 + 2 \cot \theta \eta^2_3 + 2 \eta^3_{3,\phi} - \xi_{3,s} = 0, \nonumber \\& & \eta^0_{3,r} - \eta^1_{3,t} - \frac{1}{r} (\eta^0_{2,r} + \eta^1_{2,t} ) - \frac{1}{r^2} \eta^1_{1,t} \nonumber\\& & \quad + \frac{3 k}{2 r^3} \left[ \eta^0_{0,r} + (3 k -2) \eta^1_{0,t} \right] = 0, \nonumber  \\& & \eta^0_{3,\theta} - r^2 \eta^2_{3,t} - \frac{1}{r} \eta^0_{2,\theta} + \frac{3 k}{2 r^3} \eta^0_{0,\theta}= 0, \nonumber \\& & \eta^0_{3,\phi} - r^2 \sin^2 \theta \eta^3_{3,t} - \frac{1}{r} \eta^0_{2,\phi} + \frac{3 k}{2 r^3} \eta^0_{0,\phi}= 0, \nonumber \\& & \eta^1_{3,\theta} + r^2 \eta^3_{2,r} + \frac{1}{r} \eta^1_{2,\theta} + \frac{1}{r^2} \eta^1_{1,\theta} +  \frac{(2- 3k)}{2 r^3} \eta^1_{0,\theta} = 0, \nonumber \\& & \eta^1_{3,\phi} + r^2 \sin^2 \theta \eta^3_{3,r} + \frac{1}{r} \eta^1_{2,\phi} +  \frac{1}{r^2} \eta^1_{1,\phi}  + \frac{(2- 3k)}{2r^3} \eta^1_{0,\phi} = 0, \nonumber\\
& &  V_{,t} \eta^0_3 + V_{,r} \eta^1_3 + V_{,\theta} \eta^2_3 + V_{,\phi} \eta^3_3 + V \xi_{3,s} + A_{3,s} = 0.  \nonumber
\end{eqnarray}
We simultaneously solved exact, the first-order, the second-order and the third-order ANGS equations (\ref{angs-eq1}), (\ref{angs-eq2}), (\ref{angs-eq3}) and (\ref{angs-eq4}) for the constant potential, e.g. $V(t,r,\theta,\phi) = V_0$. From Eqs.(\ref{angs-eq1}), (\ref{angs-eq2}), (\ref{angs-eq3}) and (\ref{angs-eq4}), we find for the exact ANGSs
\begin{eqnarray}
& & \xi_0 = a_1, \, \eta^0_0 = a_2, \, \eta^1_0 = 0, \, \eta^2_0 = -a_3 \cos \phi + a_4 \sin \phi, \quad \,\, \label{soln-eq1-1} \\ & & \eta^3_0 = \cot \theta (a_3 \sin \phi + a_4 \cos\phi) + a_5, \quad A_0 = a_6. \label{soln-eq1-2}
\end{eqnarray}
for the first-order ANGSs
\begin{eqnarray}
& & \xi_1 = b_1, \, \eta^0_1 = b_2, \, \eta^1_1 = 0, \, \eta^2_1 = -b_3 \cos \phi + b_4 \sin \phi, \quad \,\, \label{soln-eq2-1} \\ & & \eta^3_1 = \cot \theta (b_3 \sin \phi + b_4 \cos\phi) + b_5, \quad A_1 = b_6. \label{soln-eq2-2}
\end{eqnarray}
for the second-order ANGSs
\begin{eqnarray}
& & \xi_2 = c_1, \, \eta^0_2 = c_2, \, \eta^1_2 = 0, \, \eta^2_2 = -c_3 \cos \phi + c_4 \sin \phi, \quad \,\, \label{soln-eq3-1} \\ & & \eta^3_2 = \cot \theta (c_3 \sin \phi + c_4 \cos\phi) + c_5, \quad A_2 = c_6. \label{soln-eq3-2}
\end{eqnarray}
and for the third-order ANGSs
\begin{eqnarray}
& & \xi_3 = d_1 \frac{s^2}{2} + d_2 s + d_3, \label{soln-eq4-1} \\& & \eta^0_3 = \frac{s}{2} ( d_1 t + 2 d_4) + d_2 \frac{t}{2} + d_5, \label{soln-eq4-2} \\& & \eta^1_3 = \frac{r}{2} ( d_1 s + d_2), \quad \eta^2_3 = d_6 \cos \phi + d_7 \sin \phi, \label{soln-eq4-3} \\ & & \eta^3_3 = \cot\theta \left( d_6 \cos\phi  - d_7 \sin\phi \right) +  d_8, \label{soln-eq4-4} \\& & A_3 = \frac{1}{4} d_1 \left( t^2  - r^2 - 2 V_0 s^2 \right) - d_2 V_0 s + d_4 t + d_9 , \label{soln-eq4-5}
\end{eqnarray}
where $a_{\ell}, b_{\ell}, c_{\ell} (\ell=1,\ldots, 6)$ and $d_1, \ldots, d_9$ are the arbitrary constants of integration associated with the ANGSs.

It is easily seen from the above solutions that we have 5 exact, first-order and second-order ANGSs and 8 third-order ANGSs which includes also exact ones. It is known that the spacetime of Bardeen model has four KVs, i.e.,
\begin{eqnarray}
& & {\bf K}_0 = \p_t, \qquad {\bf K}_1 = \cos \phi \p_{\theta} - \cot \theta \sin \phi \p_{\phi}, \label{kv-01} \\ & &  {\bf K}_2 = \sin \phi \p_{\theta} + \cot\theta \cos\phi \p_{\phi}, \qquad  {\bf K}_3 = \p_{\phi}, \label{kv-23}
\end{eqnarray}
which correspond to the conservation of energy and angular momentum only.

For all order of the ANGSs, we point out here that there is the vector field ${\bf Y}_0 = \p_s$ which gives translation in geodetic parameter $s$ and it always exists for a geodesic Lagrangian of the type (\ref{lagr}), and remaining ones are  the \emph{four} KVs given in (\ref{kv-01}) and (\ref{kv-23}). We found that the exact,the first-order and the second-order ANGS algebras admitted by Bardeen model are \emph{five} dimensional. In addition to those of five ANGSs, there are three additional third-order ANGSs as the following:
\begin{eqnarray}
& & {\bf Y}_1 = s {\bf Y}_0 , \quad \,\, {\rm with \,\, gauge \,\, term \,\,} A_3 = t, \label{b-Y2} \\& & {\bf Y}_2 = s \p_s + \frac{1}{2} \left( t \p_t + r \p_r \right), \label{b-Y1} \\& & \quad \qquad {\rm with \,\, gauge \,\, term \,\,} A_3 = - V_0 s, \nonumber  \\& & {\bf Y}_{3} = s \left( s \p_s + t \p_t + r \p_r \right),\label{b-Y3} \\& & \quad \qquad {\rm with \, gauge \, term \,\,} A_3 = \frac{1}{2} \left( t^2 -r^2 - 2 V_0 s^2 \right). \nonumber
\end{eqnarray}
Thus, it is seen that for the Bardeen model the conservation laws of linear momentum and spin angular momentum in the third-order approximation are lost. In the third-order approximate symmetry case, we compute the ANGSs of perturbed geodesic Lagrangian for the Bardeen metric with non-constant potential. In order to determine the ANGSs associated with a given non-constant potential $V(t,r,\theta,\phi)$, one has to use the new ANGS Eqs.(\ref{angs-eq1}), (\ref{angs-eq2}), (\ref{angs-eq3}) and (\ref{angs-eq4}). We have collected our results in Table \ref{table-1}.

Now, we intend to recover the lost conservation laws. Therefore, it will be considered the lower order than third-order, i.e. the second-order, approximate symmetry of the Bardeen model. Thus we will retain the second-order terms in $\epsilon$ and neglect $O(\epsilon^3)$ in the geodesic Lagrangian (\ref{Bardeen-lagr}) for the Bardeen model. It is noted that the metric of Bardeen model as the second-order perturbed spacetime reduces to the second-order perturbed Schwarzschild solution. So we will use the new set of determining equations up to second-order to check whether the conservation laws are lost or not. Solving the exact, first-order and second-order ANGS equations (\ref{angs-eq1}), (\ref{angs-eq2}) and (\ref{angs-eq3}) for the constant potential it follows the same components given by (\ref{soln-eq1-1})-(\ref{soln-eq1-2}) and (\ref{soln-eq2-1})-(\ref{soln-eq2-2}) for the exact and first-order ANGSs, respectively. But for the second-order ANGSs we have the following components
\begin{eqnarray}
& & \xi_2 = c_1 \frac{s^2}{2} + c_2 s + c_3, \label{soln-eq3-1} \\& & \eta^0_2 = \frac{t}{2} ( c_1 s + c_2) - r \sin \theta ( c_4 \cos \phi - c_5 \sin \phi) \nonumber \\& & \qquad \quad - c_6 r \cos \theta + c_7 s + c_8, \label{soln-eq3-2} \\& & \eta^1_2 = \frac{r}{2} ( c_1 s + c_2) + t \sin \theta ( - c_4 \cos \phi + c_5 \sin \phi ) \nonumber \\& &  \quad  - \cos \theta (c_6 t + c_9 s + c_{10}) - \sin \theta \Big{[} s ( c_{11} \sin \phi - c_{12}  \cos \phi ) \nonumber \\& &  \qquad \qquad \qquad + c_{13} \cos\phi - c_{14} \sin \phi \Big{]}, \label{soln-eq3-3} \\& &  \eta^2_2 = \frac{\cos \theta}{r} \Big{[} t ( - c_4 \cos \phi + c_5 \sin \phi)  \nonumber \\& & \qquad \qquad  - s ( c_{11} \sin\phi + c_{12} \cos\phi ) - c_{13} \cos\phi  + c_{14} \sin\phi \Big{]}  \nonumber \\& &  \qquad + \frac{\sin\theta}{r} \left( c_6 t + c_9 s + c_{10} \right) - c_{15} \cos \phi + c_{16} \sin \phi, \label{soln-eq3-4} \\ & & \eta^3_2 = \frac{1}{r \sin\theta} \Big{[} t ( c_4 \sin\phi + c_5 \cos\phi)  \nonumber \\& & \qquad \quad + s (- c_{11} \cos\phi + c_{12} \sin\phi ) + c_{13} \sin\phi + c_{14} \cos\phi \Big{]} \nonumber \\& &  \qquad   + \cot\theta \left( c_{15} \sin\phi + c_{16} \cos\phi \right) +  c_{17}, \label{soln-eq3-5} \\& & A_2 = \frac{1}{4} c_1 \left( t^2  - r^2 - 2 V_0 s^2 \right) - c_2 V_0 s + c_7 t + c_9 r \cos\theta \nonumber \\& & \qquad \quad + r \sin\theta ( c_{11} \sin\phi + c_{12} \cos\phi ) + c_{18}, \label{soln-eq3-6}
\end{eqnarray}
where $c_1,\ldots,c_{18}$ are constants of integration. It is immediately seen from these solutions that the ``lost'' symmetries for the Bardeen metric are appeared in the second-order ANGS solutions, and the second-order ANGS generators are the four KVs (\ref{kv-01})-(\ref{kv-23}), and ${\bf Y}_{0},\ldots, {\bf Y}_3$ and ${\bf Z}_{1},\ldots, {\bf Z}_9$ given by (\ref{Y0})-(\ref{Z9}) in Appendix, where for the second-order ANGSs of Bardeen model, the gauge functions are second-order, ${\bf Y}_2$ has a gauge function such as $A_2 = - V_0 s$ and the gauge function of ${\bf Y}_3$ is also of the form $A_2 = \frac{1}{2}( t^2 -r^2 -2 V_0 s^2)$. Hence, we have recovered seventeen ANGSs in second-order approximate case of the perturbed Lagrangian for Bardeen model as it is expected. We conclude that the first-order ANGSs that have been recovered by the first-order perturbed Schwarzschild metric are the second-order ANGSs for the Bardeen metric, which is also equivalent to the second-order perturbed Schwarzschild metric. The ANGSs and corresponding gauge functions of the  second-order perturbed Schwarzschild solution are also computed for some non-constant potentials, and the results of the calculations are same as the ones shown in the Table \ref{table-1} of Ref.\cite{camci2014c}, but there is only one difference that the gauge function $A_1$ of first-order perturbation should be replaced by the gauge function $A_2$ of second-order perturbation in Table \ref{table-1} of Ref.\cite{camci2014c}.

\begin{table*}[!ht]
\centering
\caption{Third-order Approximate Noether gauge symmetries and associated gauge functions $A_0, A_1, A_2$ and $A_3$ of the Bardeen metric for some non-constant potentials $V(t,r,\theta,\phi)$. Here $V_0$ is a non-zero constant.} \label{table-1}
\resizebox{17cm}{!} {
\begin{tabular}{@{\extracolsep{\fill}}llllll@{}}
\hline
Potential & ANGSs & $A_0$ & $A_1$ & $A_2$ & $A_3$    \\
\hline
$V =\frac{V_0}{r}$  & ${\bf K}_0, \, {\bf K}_1, \, {\bf K}_2, \, {\bf K}_3, \, {\bf Y}_0$& $ 0 $ & $ 0 $ & $ 0 $ & $ 0 $ \\  & ${\bf Y}_1 $ &  $ 0$ &  $ 0$ &  $ 0$ &  $ t$ \\
    \hline
$V =\frac{V_0}{r^2}$  & ${\bf K}_0, \, {\bf K}_1, \, {\bf K}_2, \, {\bf K}_3, \, {\bf Y}_0, \, {\bf Y}_{2}$& $ 0 $ & $ 0 $ & $ 0 $ & $ 0 $ \\  & ${\bf Y}_1, \, {\bf Y}_{3} $ & $ 0$ & $ 0 $ & $ 0 $ & $ t, \, \frac{1}{2} (t^2 - r^2)$ \\[1mm]
    \hline
$V = \frac{V_0}{r \sin \theta}$ & ${\bf K}_0, \, {\bf K}_3, \, {\bf Y}_0$ & $0$ & $ 0$ & $ 0$ & $ 0$ \\
      & ${\bf Y}_1 $ & $ 0$  & $ 0$ & $ 0$ & $ t$ \\
    \hline
$V = \frac{V_0}{r^2 \sin \theta}$ & ${\bf K}_0, \, {\bf K}_3, \, {\bf Y}_0, \, {\bf Y}_2$ & $0$ & $ 0$ & $ 0$ & $ 0$ \\
      & ${\bf Y}_1, \, {\bf Y}_3$ & $ 0$ & $ 0$ & $ 0$ & $ t, \, \frac{1}{2} (t^2 - r^2)$  \\[1mm]
    \hline
$V = V_0 \, r^2 $ & $ {\bf K}_0,\, {\bf K}_1, \, {\bf K}_2, \, {\bf K}_3, \, {\bf Y}_0 $ & $ 0$  & $ 0$ & $ 0$ & $ 0$ \\
& ${\bf Y}_1$ & $ 0$ & $ 0$ & $ 0$ & $ t$ \\
    \hline
$V = V_0 \, t^2$ & ${\bf K}_1,\, {\bf K}_2, \, {\bf K}_3, \, {\bf Y}_0 $ & $ 0 $ & $ 0 $ & $ 0 $ & $ 0 $
         \\
         & $ \sin (\sqrt{2 V_0} s) {\bf K}_0, \,\, \cos (\sqrt{2 V_0} s) {\bf K}_0 $ & $ 0 $ & $ 0 $ & $ 0 $ & $ \sqrt{2 V_0} \cos (\sqrt{2 V_0} s), \,\, -\sqrt{2 V_0} \sin (\sqrt{2 V_0} s) $ \\[1mm]
    \hline
$V= \frac{V_0}{8} (r^2 - t^2) $ & $ {\bf K}_1,\, {\bf K}_2, \, {\bf K}_3, \, {\bf Y}_0$ & $ 0 $ &  $ 0 $ &  $ 0 $ &  $ 0 $ \\
   & $ e^{V_0 s}\left[ (1- v_0 s) {\bf Y}_0 + V_0 {\bf Y}_2 \right] , \, e^{V_0 s}\left[ (1 + v_0 s) {\bf Y}_0 - V_0 {\bf Y}_2 \right], \, e^{V_0 s/2} {\bf Y}_{0}, \, e^{-V_0 s/2} {\bf Y}_{0} $ & $ 0 $ & $ 0 $ & $ 0 $ & $ -\frac{V_0^2}{4} (t^2 -r^2) e^{V_0 s}, \, -\frac{V_0^2}{4} (t^2 -r^2) e^{-V_0 s}, - \frac{V_0}{2} t e^{-V_0 s /2}, \, \frac{V_0}{2} t e^{V_0 s /2}  $  \\[1mm]
    \hline
$V= V_0 e^{\lambda \phi}$ & $ {\bf K}_0, \, {\bf Y}_0, \, {\bf Y}_2 - \frac{1}{\lambda} {\bf K}_3 \, (\lambda \neq 0) $ & $ 0 $ & $ 0 $ & $ 0 $ & $ 0 $ \\
   & $ {\bf Y}_1 $ & $ 0 $ & $ 0 $ & $ 0 $ & $ t $ \\
	\hline
$V = V_0 \phi$ & ${\bf K}_0, \, {\bf Y}_0$ & $ 0 $ & $ 0 $ & $ 0 $ & $ 0 $ \\
   & $ {\bf K}_3, \,\, {\bf Y}_1 $ & $ - V_0 s, 0 $ & $ - V_0 s, 0 $ & $ -V_0 s, 0 $ & $ -V_0 s, \,\, t $ \\
	\hline

$V = V_0 t$ & ${\bf K}_1, \, {\bf K}_2, \, {\bf K}_3, \, {\bf Y}_0$ & $ 0 $ & $ 0 $ & $ 0 $ & $ 0 $ \\
   & $ {\bf K}_0, \,\,\, {\bf Y}_2 - \frac{3 V_0 s^2}{4} {\bf Y}_0, \,\,\, {\bf Y}_3 - \frac{V_0 s^3}{4} {\bf Y}_0$ & $ - V_0 s, 0, 0 $ & $ - V_0 s, 0, 0 $ & $ - V_0 s, 0, 0 $ & $ - V_0 s, \,\,\, \frac{V_0 s }{2} (\frac{s^2}{2} - 3 t), \, \,\, \frac{1}{2} (t^2 - r^2) + \frac{V_0 s^2 }{2} (\frac{V_0 s^2}{4} - 3 t)  $ \\[1mm]
	\hline

\end{tabular}
}
\end{table*}

\section{Conclusions and Discussions}
\label{sec:4}
In this paper, we studied the geometric nature of ANGSs up to third-order (in $\epsilon$) for the geodesic Lagrangian of spacetimes. We presented that the computation of third-order ANGSs of the perturbed first-order geodesic Lagrangian (\ref{p-lagr}) is reduced to the problem of finding solutions for the new geometrical set of exact differential conditions (\ref{aneq-1-1})-(\ref{aneq-1-3}), first-order differential conditions (\ref{aneq-2-1})-(\ref{aneq-2-3}), second-order differential conditions (\ref{aneq-3-1})-(\ref{aneq-3-3}) and third-order differential conditions (\ref{aneq-4-1})-(\ref{aneq-4-3})  which are involve the exact, first-order, second-order and third-order ANGSs and the potential function defining the dynamical system. For the spacetime of Bardeen model, this new method of finding ANGSs is able to get an ANGS with non-zero potential.

It is pointed out in Ref.\cite{sharif2011} that one cannot recognize the parameter $e$ of the Bardeen spacetime as the Coulomb charge $Q$ of the RN spacetime. We note that the Bardeen model with magnetic charge $e$ is the analogue of the RN spacetime with electric charge $Q$. For the Bardeen model, although  there is no term of $g_{tt}$ quadratic in $\epsilon$, $g_{rr}$ includes a quadratic term (see Eqs.(\ref{Bardeen-exp1}) and (\ref{Bardeen-exp2})). Therefore, getting the lost symmetries, it is sufficient to calculate ANGSs of the RN spacetime for second-order term of perturbed Lagrangian, but this may not true for the Bardeen model in which it is needed the calculation of ANGSs up to third-order term in the perturbed Lagrangian.

Applying our new method to the Bardeen model up to third-order approximation, we found that the number of exact, first-order and second-order ANGSs is \emph{five}. The number of third-order ANGSs was found to be \emph{eight} which was not seventeen as it is expected. That is, in the third-order approximate symmetry case, the linear and spin angular momentum conservation laws for the Bardeen spacetime are lost. Retaining only terms of first and second-order in metric coefficients $g_{tt}$ and $g_{rr}$ given in (\ref{Bardeen-exp1}) and (\ref{Bardeen-exp2}), and neglecting $O(\epsilon^3)$, it reduces to the second-order perturbed spacetime for the Schwarzschild solution. Thus, for the Bardeen model as the second-order perturbed spacetime,  we reconsider our new ANGS equations to check whether those of lost conservation laws can be recovered. Then we find that for the Bardeen model all the lost conservation laws are recovered in the second-order approximation.

It would be interesting to analyse the other special spacetimes according to the ANGSs in which one can find the answer of the question that ``what is the order of ANGSs to give all conservation laws''. This issue will be a subject of research in future.

\section*{Appendix: ANGSs for Schwarzschild spacetime}

The number of ANGS generators of the first order (in $\epsilon$) geodesic Lagrangian for the Schwarzschild spacetime with vanishing potential are seventeen \cite{hussain2009} which are the four KVs given in (\ref{kv-01}) and (\ref{kv-23}), and
\begin{eqnarray}
& & {\bf Y}_0 = \p_s, \label{Y0} \\& & {\bf Y}_1 = s {\bf Y}_0 , \,\,\, {\rm with \,\, gauge \,\, term \,\,} A_1 = t, \label{Y1} \\& & {\bf Y}_2 = s \p_s + \frac{1}{2} \left( t \p_t + r \p_r \right), \label{Y2} \\& & {\bf Y}_3 = s \left( s \p_s + t \p_t + r \p_r \right),  \label{Y3}\\& & \qquad \quad {\rm with \, gauge \, term \,\,} A_1 = \frac{1}{2} \left( t^2 -r^2 \right), \nonumber
\\& & {\bf Z}_1 = \sin \theta \cos\phi \p_r  +\frac{\cos \theta \cos \phi}{r} \p_{\theta} - \frac{\csc\theta \sin\phi}{r} \p_{\phi}, \label{Z1} \\& &  {\bf Z}_2 = \sin \theta \sin\phi \p_r  + \frac{\cos \theta \sin \phi}{r} \p_{\theta} + \frac{\csc\theta \cos\phi}{r} \p_{\phi}, \label{Z2} \\& & {\bf Z}_3 = \cos\theta \p_r - \frac{\sin\theta}{r} \p_{\theta}, \label{Z3} \\& & {\bf Z}_4 = r \sin\theta \cos\phi \p_t + t {\bf Z}_1, \label{Z4} \\& & {\bf Z}_5 = r \sin\theta \sin\phi \p_t + t {\bf Z}_2, \label{Z5} \\& & {\bf Z}_6 = r \cos\theta \p_t + t {\bf Z}_3, \label{Z6} \\& & {\bf Z}_{7} = s {\bf Z}_1, \,\,\, {\rm with \,\, gauge \,\, term \,\,} A_1 = -r \sin\theta \cos\phi, \quad \label{Z7} \\& & {\bf Z}_{8} = s {\bf Z}_2, \,\, {\rm with \,\, gauge \,\, term \,\,} A_1 = -r \sin\theta \sin\phi, \label{Z8} \\& & {\bf Z}_{9} = s {\bf Z}_3, \,\, {\rm with \,\, gauge \,\, term \,\,} A_1 = -r \sin\theta, \label{Z9}
\end{eqnarray}
where the generators ${\bf Z}_1, {\bf Z}_2, {\bf Z}_3$ provide the conservation of linear momentum, and ${\bf Z}_4, {\bf Z}_5, {\bf Z}_6$ yield the conservation of spin angular momentum. Using the generator ${\bf Y}_2$ one can write $s = t^2$ or $s = r^2$ which gives rise to interpretation the four KVs ${\bf K}_0, {\bf K}_1, {\bf K}_2, {\bf K}_3$, and eleven generators ${\bf Y}_0, {\bf Y}_1, {\bf Z}_1$, $\ldots, {\bf Z}_9$ are CKVs with conformal factor $\psi = (c_1 t^2 + c_2)/2$. It is very well known fact that the Lie algebra of the CKVs for a conformally flat spacetime is 15 dimensional.

\section*{Acknowledgements}

This work was supported by Akdeniz University, Scientific Research Projects Unit (BAP).


\begin{thebibliography}{}

\bibitem{ford} L. Ferrarese and H. Ford, Space Sci. Rev. {\bf 116}, 523 (2005)

\bibitem{hawking} S. W. Hawking, G. F. R. Ellis, \textit{The Large Scale Structure of Space-time}, (Cambridge University Press, Cambridge, 1973)

\bibitem{bardeen}  J. M. Bardeen, in Proceedings of GR5, Tbilisi, U.S.S.R, p.174 (1968)

\bibitem{garcia} E. Ayon-Beato, A. Garcia, Phys. Lett. B {\bf 493}, 149
(2000).

\bibitem{katzin} G. H. Katzin, J. Levine and W. R. Davis, J. Math. Phys. {\bf 10}, 617 (1969)

\bibitem{stephani} H. Stephani, \textit{Differential Equations: Their solution using symmetries}, p.99, (Cambridge University Press, Cambridge, 1989)

\bibitem{ibrahim} N. H. Ibragimov, \textit{``CRC Handbook of Lie group analysis of differential equations: Symmetries, exact solutions and conservation laws,''}, (CRC Press, Boca Raton, 1994)

\bibitem{capo00} S. Capozziello, G. Lambiase,  Gen. Relativ. Gravit. {\bf 32},  673 (2000)

\bibitem{camci2} U. Camci, Y. Kucukakca, Phys. Rev. D {\bf 76}, 084023 (2007)

\bibitem{camci3} Y. Kucukakca, U. Camci, I. Semiz, Gen. Relativ. Gravit. {\bf 44}, 1893 (2012)

\bibitem{feroze1} T. Feroze, F. M. Mahomed, A. Qadir, Nonlinear Dyn. {\bf 45}, 65 (2006)

\bibitem{tsamparlis} M. Tsamparlis, A. Paliathanasis, Gen. Relativ. Gravit.{\bf 42}, 2957 (2010)

\bibitem{camci1} Y. Kucukakca, U. Camci,  Astrophys. Space Sci. {\bf 338}, 211 (2011)

\bibitem{ibrar} I. Hussain, M. Jamil, F. M. Mahomed, Astrophys. Space Sci. {\bf 337}, 373 (2011)

\bibitem{feroze2} F. Ali,  T. Feroze, Int. J. Theor. Phys. {\bf 52}, 3329 (2013)

\bibitem{camci2014a} U. Camci, JCAP \ {\bf 07}, 002 (2014)

\bibitem{camci2014b} U. Camci, A. Yildirim, Phys. Scr. {\bf 89}, 084003 (2014)

\bibitem{feroze3} T. Feroze, A. H. Kara, Int. J. Non-Linear Mech. {\bf 37}, 275 (2002)

\bibitem{bokhari2006} A. H. Bokhari, A. H. Kara, A. R. Kashif, F. D. Zaman, Int. J. Theor. Phys. {\bf 45}, 1063 (2006)

\bibitem{kara2008} A. H. Kara, F. M. Mahomed, A. Qadir, Nonlinear Dyn. {\bf 51}, 183 (2008)

\bibitem{hussain2007} I. Hussain, F. M. Mahomed, A. Qadir, SIGMA \ {\bf 3}, 115 (2007)

\bibitem{hussain2009} I. Hussain, F. M. Mahomed, A. Qadir,  Gen. Relativ. Gravit. {\bf 41}, 2399 (2009)

\bibitem{sharif2012} M. Sharif, S. Waheed, Braz. J. Phys. {\bf 42}, 219 (2012)

\bibitem{wang} S. Zhou, J. Chen, Y. Wang, Int. J. Mod. Phys. D {\bf 21}, 1250077 (2012)

\bibitem{sharif2011} M. Sharif, S. Waheed,  Phys. Scr. {\bf 83}, 015014 (2011)

\bibitem{camci2014c} U. Camci, to appear in Gen. Relativ. Gravit. (2014) 





\end{thebibliography}
\end{document}